\documentclass[epj]{webofc}
\usepackage[varg]{txfonts}   

\usepackage{upgreek}
\wocname{EPJ Web of Conferences}
\woctitle{ICNFP 2016}
\begin{document}
%
\selectlanguage{english}
\title{Luminescence of water or ice as a new detection method for magnetic monopoles}
%
%

\author{Anna Obertacke Pollmann\inst{1}\fnsep\thanks{\email{anna.pollmann@uni-wuppertal.de}} for the IceCube Collaboration\footnote{\protect\url{http://icecube.wisc.edu}}
}

\institute{Dept. of Physics, University of Wuppertal, D-42119 Wuppertal, Germany 
}

\abstract{%
  Cosmic ray detectors use air as a radiator for luminescence. 
  In water and ice, Cherenkov light is the dominant light producing mechanism when the particle's velocity exceeds the Cherenkov threshold, approximately three quarters of the speed of light in vacuum.
  
  Luminescence is produced by highly ionizing particles passing through matter due to the electronic excitation of the surrounding molecules. The observables of luminescence, such as the wavelength spectrum and decay times, are highly dependent on the properties of the medium, in particular, temperature and purity. The results for the light yield of luminescence of previous measurements vary by two orders of magnitude.
  
  It will be shown that even for the lowest measured light yield, luminescence is an important signature of highly ionizing particles below the Cherenkov threshold. These could be magnetic monopoles or other massive and highly ionizing exotic particles. With the highest observed efficiencies, luminescence may even contribute significantly to the light output of standard model particles such as the PeV IceCube neutrinos.
  
  We present analysis techniques to use luminescence in neutrino telescopes and discuss experimental setups to measure the light yield of luminescence for the particular conditions in neutrino detectors.
}
\maketitle
%


\section{Magnetic monopoles}
\label{sec:monopoles}

\begin{figure}[t]
\centering
\includegraphics[width=\textwidth,clip]{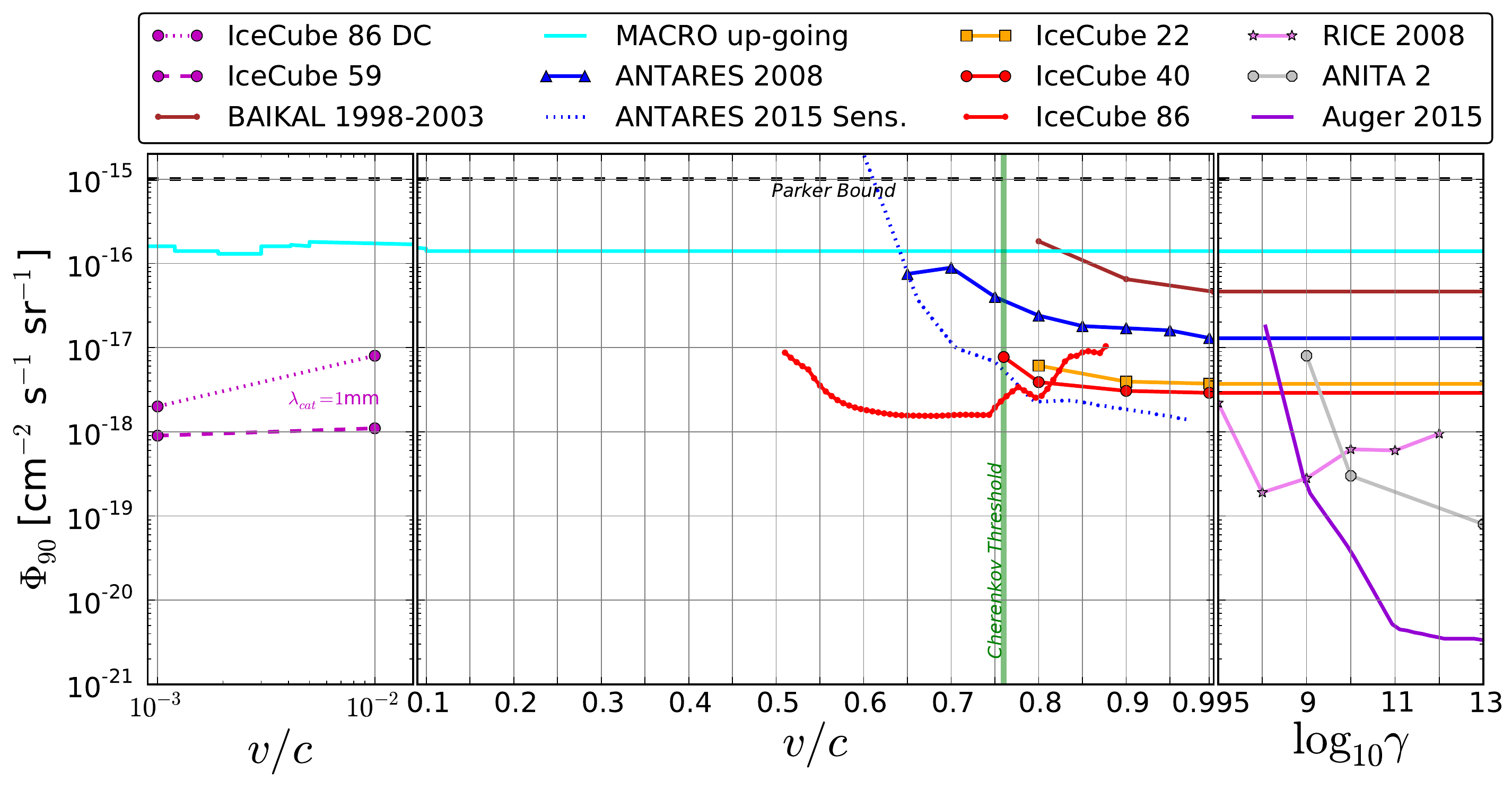}
\caption{Current upper limits on the flux of magnetic monopoles in dependence on the 	speed of monopoles at the detector \cite{Rice08,Baikal08,Anita11,Antares12,Christy13,Schoenen14,Antares15,Auger15,ICMonopoles16}. Best limits for cosmic monopoles are achieved by general purpose detectors only. However, low relativistic speeds, from $0.1\,c$ to $0.5\,c$, are not covered by them yet so that the latest limit in this region is almost two orders of magnitude higher than for all other speed ranges}
\label{fig:limits}       
\end{figure}

The existence of magnetic monopoles, particles carrying a single magnetic charge, is motivated by various theories which extend the Standard Model of particles, such as Grand Unified Theories (GUTs), String theory, Kaluza-Klein, and M-Theory \cite{Shnir05}. The elementary magnetic charge $g_D$ can be found by basic considerations to be dependent on the elementary electric charge 
$g_D = e/2 \alpha \approx 68.5 \, e$ where $e$ is the elementary electric charge and $\alpha$ is the fine structure constant \cite{Dirac1931}. Other parameters of magnetic monopoles, such as their mass and predicted flux, highly depend on the details of the models regarding particle creation. In most theories the mass of magnetic monopoles is in the order of 
$10^7\, \mathrm{GeV}/c^2$ to $10^{13}\, \mathrm{GeV}/c^2$ (intermediate mass monopoles), or 
$10^{13}\, \mathrm{GeV}/c^2$ to $10^{19}\, \mathrm{GeV}/c^2$ (GUT /some SUSY monopoles).
The  mass is so high that monopoles cannot be created in any foreseeable particle accelerator but rather shortly after Big Bang through (intermediate steps of) gauge symmetry breaking \cite{Rajantie12}. The proposed inflationary phase of the early universe leads to a dilution of the monopole density. Current experimental limits on the monopole flux are stronger than theoretical bounds. 
An overview of the most recent experimental limits as a function of the monopole speed at the detectors is given in Fig. \ref{fig:limits}. 
Monopoles at relativistic speeds are expected to have intermediate masses since only in this mass range they can be sufficiently accelerated in cosmic magnetic fields \cite{Wick03}.

Special purpose detectors for monopoles use induction, visible damage of plastic targets, and time of flight measurement such as MACRO or MoEDAL \cite{Macro02,Moedal16}. General purpose particle detectors are used for monopole searches if they have larger effective areas, leading to an increased sensitivity compared with special purpose detectors. This condition is fulfilled by all large scale neutrino telescopes, Baikal, ANTARES, and IceCube as shown in Fig. \ref{fig:limits}.


\section{Monopole detection with IceCube}
\label{sec:icecube}

IceCube is a cubic-kilometer neutrino detector installed in the ice at the geographic South Pole \cite{IceCube06} between depths of 1450 m and 2450 m. It comprises 86 strings each with 60 digital optical modules (DOMs). Neutrino reconstruction relies on the optical detection of Cherenkov radiation emitted by secondary particles produced in neutrino interactions in the surrounding ice or the nearby bedrock.
The DeepCore sub-array, as defined in this work, includes 8 more densely instrumented strings optimized for low energies plus 12 adjacent standard strings.

The detector is triggered when a DOM and its nearest or next-to-nearest DOMs record a hit within a $1\,\mu \textrm{s}$ window. Then all hits in the detector within a window of $10\, \mu \textrm{s}$ 
will be readout and combined into one event \cite{IceCubeDAQ09}. A series of data filters are run on-site in order to select potentially interesting events for further analysis, reducing 
the amount of data to be transferred 
via satellite.

Magnetic monopoles faster than the Cherenkov threshold in ice, $v_C \approx 0.76\,c$, emit Cherenkov light analogous to electrically charged particles. In addition monopoles ionize the surrounding matter, and most of the knocked-off $\delta$-electrons are ejected with energies sufficient to produce Cherenkov light. This mechanism is called indirect Cherenkov light emission and produces light down to a monopole velocity of $ \gtrsim 0.45\,c$. 
At ultra-relativisitc speeds radiative losses start to dominate the light production. In water these are dominantly photonuclear reactions.

\begin{figure}[t]
	\begin{minipage}{0.64\linewidth}
		\centering
		\includegraphics[width=0.9\textwidth,clip]{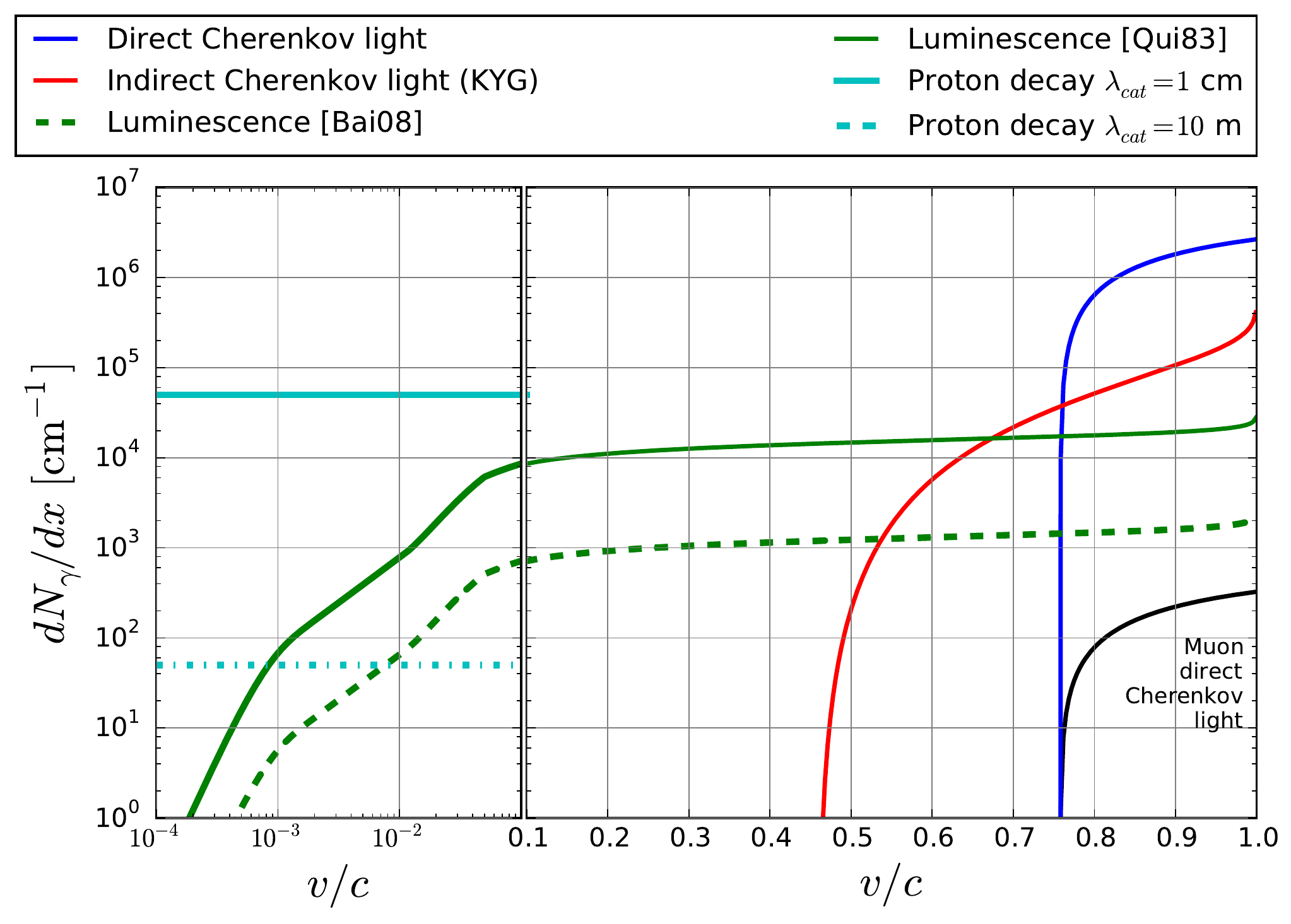}
		\caption{
			Light yield for the different detection mechanisms of magnetic monopoles in neutrino telescopes (here assuming the refraction index of ice). For comparison the direct Cherenkov light emitted by a bare muon is shown.
			The luminescence light yield is calculated by multiplying the luminescence efficiency with the energy loss of magnetic monopoles taken from Ref. \cite{Derkaoui1998}. Quenching effects are not taken into account because they are not measured for water yet
		}
		\label{fig:lightyield}       
	\end{minipage}
	\hfill
	\begin{minipage}{0.34\linewidth}
		\centering
		\vspace{0.2cm}
		\includegraphics[width=0.9\textwidth,clip]{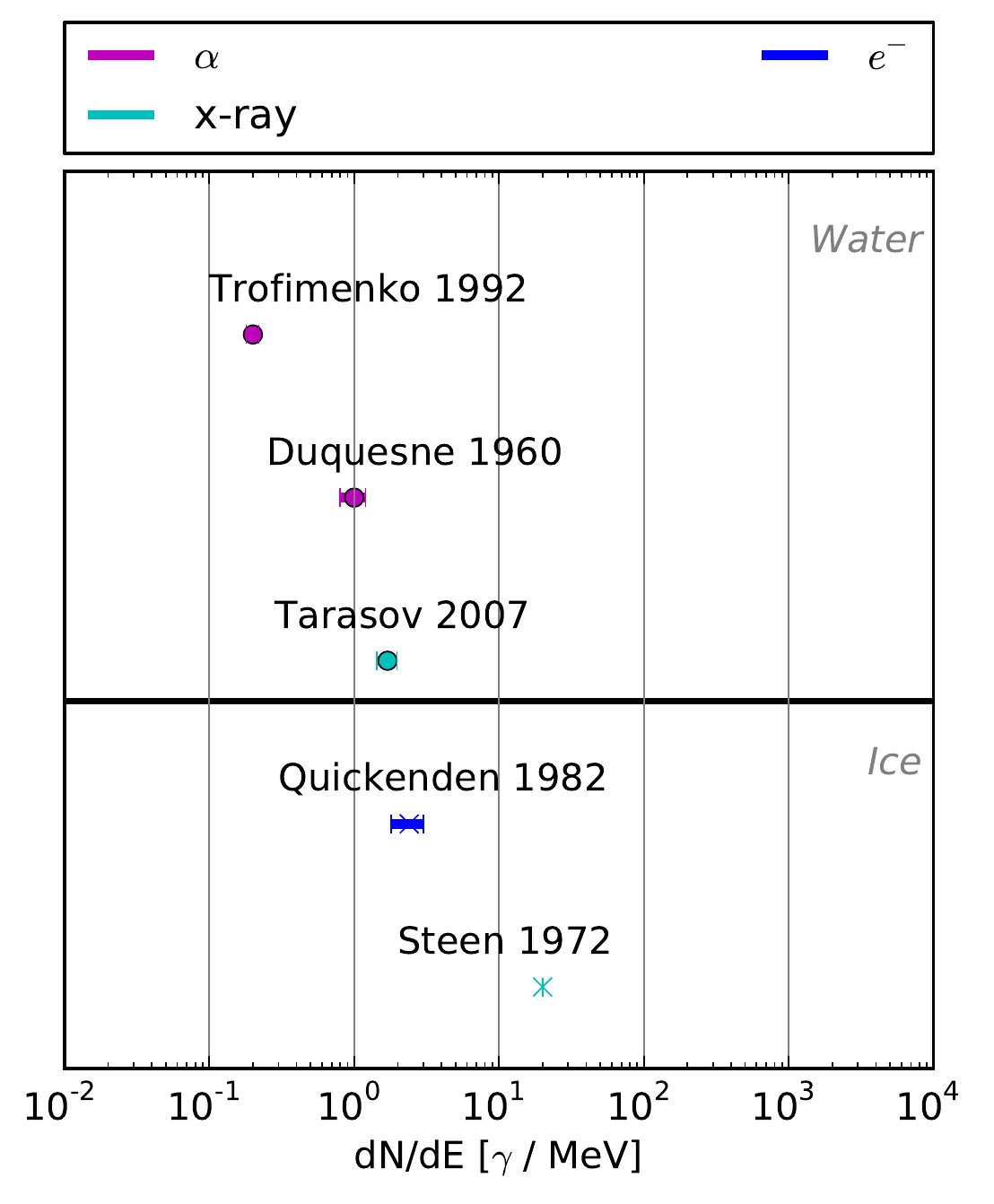}
		\caption{Overview of luminescence measurements at different phases of matter and using different irradiations, taken from Refs. \cite{Trofimenko92,Duquesne60, Tarasov2007, Quickenden82,Steen72}.  Further measurements are summarized in Ref. \cite{Quickenden82}
		}
		\label{fig:efficiency}       
	\end{minipage}
\end{figure}

Some theories predict the catalysis of proton decay by magnetic monopoles \cite{Rubakov88}. The decay products start particle cascades which are visible through Cherenkov radiation of the particles. If realized in nature, the Rubakov-Callan mechanism is the dominating light production mechanism at low monopole speeds.

An overview over the light yield of magnetic monopoles from different production mechanisms in dependence on the particle speed is given in Fig. \ref{fig:lightyield}.
Low relativistic speeds, from $0.1\,c$ to $0.45\,c$, are not covered by the described mechanisms, but may be covered by luminescence.


\section{Luminescence of water}
\label{sec:water_lumi}

Luminescence is defined in this text as the excitation of transparent media by ionizing radiation resulting in visible light.
It was first concluded in 1953 that the luminescence, measured in water, originates from the water molecules itself and not from impurities \cite{Belche53}. Subsequently the effect was investigated in regard of the luminescence efficiency, lifetime of excited states, and wavelength spectrum to find the underlying mechanisms. The early results are summarized in Ref. \cite{Quickenden82}. Later measurements of the luminescence efficiency, or measurements not covered in this reference, are summarized in Fig. \ref{fig:efficiency}.

The spectrum of luminescence peaks at visible wavelengths around $400\,\mathrm{nm}$ \cite{Quickenden82}. The measurements of the lifetimes of excited states are inconclusive. It was debated that the luminescence originates from transitions of excited OH$^-$ or H$_3$O$^+$ molecules \cite{Quickenden86}.

In Ref. \cite{Quickenden82} it is concluded that impurities in water enhance the luminescence efficiency.
Also it is shown that the luminescence efficiency is highly dependent on the temperature \cite{Quickenden91}. 


\section{Luminescence induced by magnetic monopoles}

\begin{figure}[t]
	\centering
	\includegraphics[width=0.4\textwidth,clip]{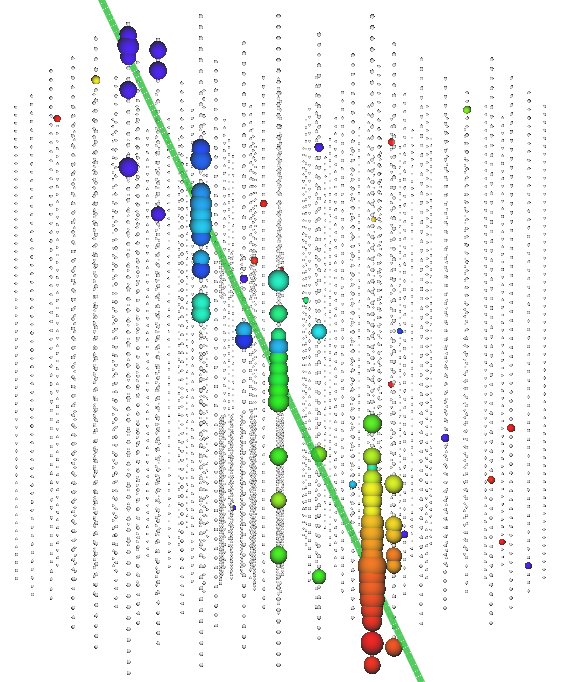}
	\hfill
	\includegraphics[width=0.4\textwidth,clip]{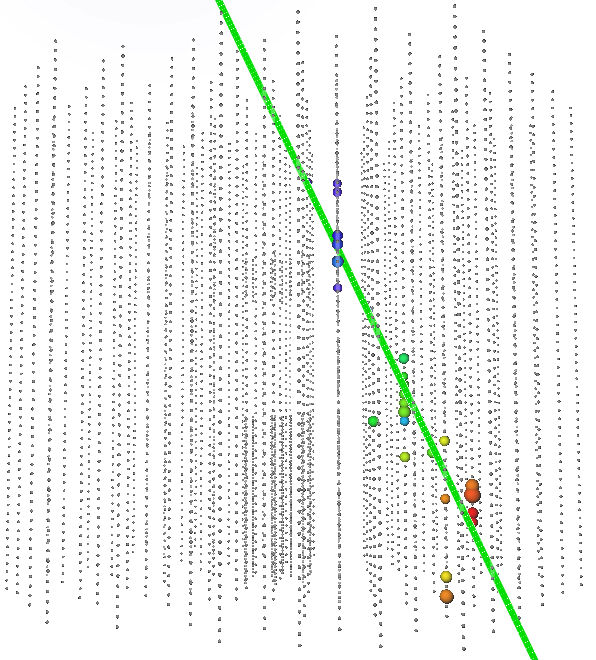}
	\caption{Event view of simulated magnetic monopoles. The gray dots represent the 5160 digital optical modules (DOMs) of IceCube. When a DOM has seen light it is colored from red (early) to blue (late) regarding hit time. The more light was recorded, the larger the radius of a DOM is enhanced. Both events develop up-going from the bottom right to the top left.
    \textbf{Left:} The simulated speed is $v=0.52\,c$. The light originates from indirect Cherenkov radiation only. When luminescence light is added according to the light yield given in Fig. \ref{fig:lightyield}, the signature would be even brighter.
	\textbf{Right:} The simulated speed is $v=0.3\,c$. The light originates from luminescence assuming a conservative efficiency of $dN_{\gamma}/dE=0.2\,\gamma/\mathrm{MeV}$ , as given in Ref. \cite{Baikal08}, integrated over wavelengths from $300\,\mathrm{nm}$ to $600\,\mathrm{nm}$ and with an assumed lifetime of $500\,\mathrm{ns}$}
	\label{fig:simulated_monopoles}       
\end{figure}

For Fig. \ref{fig:lightyield} the light yield of monopoles through luminescence is estimated using the measurement results $dN_{\gamma}/dE=0.2\,\gamma/\mathrm{MeV}$, given in Ref. \cite{Baikal08}, and $dN_{\gamma}/dE=2.4\,\gamma/\mathrm{MeV}$, given in Ref. \cite{Quickenden82}. This values were chosen to include both a conservative and an optimistic value, for the actual conditions at the neutrino detectors. 
As shown, luminescence can even complement the Cherenkov light from proton decays at non relativistic speeds, with the additional advantage that it is independent of the monopole model and catalysis cross section.

Event signatures of magnetic monopoles with low relativistic speeds between $0.1\,c$ and $0.5\,c$ inducing luminescence light are simulated for the current data acquisition and trigger of IceCube. Event views are shown in Fig. \ref{fig:simulated_monopoles} and the trigger efficiency is shown in Tab. \ref{tab:trigger_filter}. Adjustments of the triggers for higher sensitivity to slow and faint events requires an upgrade in computing resources which is feasible in short term.
However, the IceCube detector in its current configuration is already capable of detecting luminescence signatures induced by magnetic monopoles at the considered speeds. 

To use luminescence of water as a particle detection method, the exact properties have to be measured. In addition, the IceCube filters have already been extended to address this additional signature.


\section{New IceCube filter for monopoles}

As shown in Fig. \ref{fig:simulated_monopoles}, the signature of low relativistic magnetic monopoles inducing luminescence is a track which develops slowly through the whole detector. The brightness of the track decreases with lower speeds.

To achieve best filter efficiencies for both, faint and bright tracks, two selection criteria were developed for the new monopole filter. 
The \textit{IceCube selection} (IC) is applied on hits recorded with all IceCube strings but the denser instrumented strings belonging to the sub-array DeepCore. 
The \textit{DeepCore selection} (DC) is applied on the DeepCore strings and the adjacent IceCube strings. 
Both selections apply cuts on the minimum number of hit DOMs (IC/DC: 6), the maximum reconstructed speed (IC: $0.8\,c$, DC: $0.6\,c$) \cite{ImprLineFit14} to reduce background from muon signatures,  and the minimal event time length\footnote{Time difference between the last and the first hit of an event. Only the first pulse per DOM is taken into account.} (IC: $4000\,\mathrm{ns}$, DC: $2750\,\mathrm{ns}$). 
In addition, the IceCube selection requires a minimal hit separation\footnote{Separation length of the center-of-gravity of the first and last quantile of hits which are within a cylinder with radius of $100\,\mathrm{m}$ around the reconstructed track.} of $250\,\mathrm{m}$ and a maximum gap\footnote{Longest gap between hits within a cylinder with radius of $100\,\mathrm{m}$ projected on the reconstructed track.} of $200\,\mathrm{m}$.
The DeepCore selection additionally requires an FWHM time\footnote{All first pulses per DOM are ordered in time. The charge distribution is then fit with a Gaussian and the full width of half maximum of this fit is taken as the FWHM time.}  of minimum $2500\,\mathrm{ns}$.

\begin{table}[ht]
  \centering
  \caption{Efficiency of the standard IceCube triggers and the new monopole filter in keeping simulated monopoles inducing luminescence. The filter efficiency is given with respect to the rate at trigger level. Efficiencies are given for the lowest (left in each column) and highest (right in each column) expected luminescence efficiencies. Below the speed $0.2\,c$ most faint events are selected by the DeepCore selection}
  \label{tab:trigger_filter}       
  \begin{tabular}{ll|cc|cc}
     \multicolumn{2}{c|}{} & \multicolumn{2}{c}{Standard trigger efficiency} & \multicolumn{2}{c}{Monopole filter efficiency} \\
     \multicolumn{2}{c|}{luminescence efficiency}  & $0.2\,\gamma / \mathrm{MeV}$ &  $2.4\,\gamma / \mathrm{MeV}$ & $0.2\,\gamma / \mathrm{MeV}$ &  $2.4\,\gamma / \mathrm{MeV}$ \\
     \hline
     monopole speed & $0.1\,c$ &  $\,\,$ 9\% & 60\% & 25\%& 45\% \\
    & $0.3\,c$ & 36\% & 68\% & 58\% & 60\% \\
    & $0.5\,c$ & 46\% & 70\% & 64\% & 71\% \\
  \end{tabular}
\end{table}

The filter efficiency, shown in Tab. \ref{tab:trigger_filter}, is up to twice the efficiency compared to the usage of the previously installed filters not targeting  monopoles. The filter rate is $30.8\,\mathrm{Hz}$ adding $1.4\,\mathrm{GB}$ to the bandwidth of about $100\,\mathrm{GB}$ which are transfered every day via satellite from the South Pole to the northern computer farms. The filter was installed in May 2016. 
The pass rate of low relativistic monopoles could be enhanced by accounting for this signature in the detector design and data acquisition systems closer to the hardware than the filters when a detector upgrade is due.


\section{Measuring the properties of luminescence}

To use luminescence as a detection method, one needs to measure the luminescence efficiency, the lifetimes of excited states, and the wavelength spectrum at the given conditions from the detectors. 
Two types of measurements are planned, one in situ and one in the laboratory, which complement each other. Since there are significant discrepancies between the results of previous measurements, it is mandatory to have reliable results of this work for further use of luminescence.

Luminescence depends on the radiation type and charge due to quenching effects. Since the luminescence of water itself is hardly studied, the dependency of quenching on the energy loss of the incident particle charge $Z$ is unknown. The planned procedure of the laboratory measurement is to mimic a single magnetic charge $g_D$ with a corresponding electric charge $Ze$ with $Z$ close to $g_D/e\approx68.5$. If a heavy ion of this charge penetrates a water probe from the detector site with speed smaller than the Cherenkov threshold and comparable energy loss, the emitted light is comparable to the light emitted by a magnetic monopole. For preparation, measurements are planned using radioactive sources and an electron gun as well as a target of ultra pure water. The latter is done to study the luminescence of pure water itself and the influence of different added substances. In addition, the temperature dependence will be measured in detail.

The standard model particles measured in neutrino telescopes also induce the emission of luminescence light. However, the contribution of luminescence to the whole light output is at most 10\% assuming the optimistic luminescence efficiency of $dN_{\gamma}/dE=2.4\,\gamma/\mathrm{MeV}$. The contribution of luminescence can be identified in the time distribution of recorded light due to the long lifetimes. 
There are two kind of events which will be used for that. 
The first type are the astrophysical neutrino signatures which have the largest deposited energy, thus the largest luminescence contribution. 
The second type are minimal ionizing vertical muons moving close and along a string of DOMs. 
This event type is abundant in IceCube and the Cherenkov light can be used as a trigger.


\section{Summary and outlook}

The feasibility, to use luminescence as a new detection method for magnetic monopoles in neutrino detectors, is studied with IceCube. 
The signature of monopoles inducing luminescence is very likely detectable, thus a search for monopoles with low relativistic speeds would be possible. 
For this purpose, a new filter was developed and installed at IceCube, to identify the corresponding event signatures and to collect the events for future analysis. 
In addition, measurements are planned 
to improve  knowledge of the properties of water and ice luminescence which allow a proper interpretation of the corresponding IceCube results.


\bibliography{bib}

\end{document}